\documentclass[aps,prresearch,reprint,twocolumn,superscriptaddress]{revtex4-2}

\setcitestyle{super}  

\usepackage[T1]{fontenc}
\usepackage[utf8]{inputenc}

\usepackage{amsmath,amsfonts,amssymb,bm}
\usepackage{physics}

\usepackage{color,graphicx}
\usepackage{graphicx}

\usepackage[citecolor=blue]{hyperref}
\usepackage{enumitem}
\usepackage{bm}

\makeatletter
\usepackage{appendix}

\usepackage{longtable}%
\usepackage{colortbl}%

\setlist[itemize]{noitemsep}
\global\long\def\mat#1{\mathsf{#1}}%

\global\long\def\ket#1{\left|#1\right\rangle }%

\global\long\def\bra#1{\left\langle #1\right|}%
 
\global\long\def\braket#1#2{\left.\left\langle #1\right|#2\right\rangle }%
\makeatother

\begin{document}
\title{Adaptive Multiset Tree Tensor Networks for Large-Scale Exciton Dynamics}

\author{Lachlan P Lindoy}
\email{lachlan.lindoy@npl.co.uk}
\affiliation{National  Physical  Laboratory,  Teddington,  TW11  0LW,  United  Kingdom}
\author{Ivan Rungger}
\affiliation{National  Physical  Laboratory,  Teddington,  TW11  0LW,  United  Kingdom}
\affiliation{Department of Computing, Security and Mathematics, Royal Holloway, University of London, Egham, TW20 0EX, United Kingdom}

\begin{abstract}
Electronic excitations coupled to vibrational environments play an important role in charge and energy transport in molecular materials. 
Multiset tree tensor networks  provide an efficient ansatz for the dynamics of such systems by allowing different electronic configurations to develop independent environmental correlations.
However, typical implementations use fixed bond dimensions, significantly limiting the accessible system size and hence their predictive capabilities for large scale systems.
To overcome these limitations, we introduce the adaptive multiset tree tensor network (AMTTN) approach, in which the bond dimensions of each configuration are adapted independently using a local energy variance based expansion scheme.
Additionally, configurations can be temporarily frozen whenever they are predicted to remain weakly populated throughout a timestep.
For the one-dimensional Holstein model, AMTTN reproduces converged exciton dynamics one to two orders of magnitude faster than fixed bond dimension multiset methods, with the dominant cost determined by the dynamically active region with only weak residual dependence on the lattice size.  
For a P3HT:PCBM heterojunction, it enables simulations of a model containing 52 oligothiophene units avoiding finite size effects occurring on dynamically relevant time scales, which lead to the previously studied 13 unit model substantially overestimating long time populations of bound charge separated states.  
Our results show that AMTTN provides a practical method for obtaining converged quantum dynamics simulations of challenging excitonic and non-adiabatic systems for which conventional multiset approaches are prohibitively expensive.
\end{abstract}

\maketitle

\section{Introduction}
The coupling of electronic excitations to nuclear vibrations plays a central role in phenomena including polaron formation, charge and energy transport, exciton dissociation and other non-adiabatic processes in molecular materials\cite{HOLSTEIN1959325,PhysRevX.10.021062,PhysRevLett.123.126601,doi:10.1021/acs.jctc.4c00751, Bakulin2016}.
Even for a single excitation accurately describing these processes is challenging when the electronic coupling, vibrational frequency, and reorganisation energy are comparable, so that there is no clear perturbative parameter.
Calculations of transport properties require lattices large enough that the propagating wavepacket does not reach the boundaries over the timescale required to resolve the relevant observables.  
Finite size reflections or boundary wrapping artifacts can significantly alter the predicted dynamics.  

Multiset tensor network methods are particularly well suited to exciton and other non-adiabatic dynamics problems.\cite{VENDRELL2011, doi:10.1142/9789812565464_0014, RAAB1999, 10.1021/jp203174e, PhysRevLett.123.126601, 10.1063/5.0301775}
By representing the environment of each electronic configuration with an independent tensor network, while retaining an explicit description of electronic state coupling, the multiset ansatz can efficiently capture the highly electronic configuration dependent environmental correlations that arise in these problems.
This flexibility can lead to substantial reductions in computational cost relative to conventional single set approaches \cite{PhysRevLett.123.126601} and enables accurate simulations of increasingly complex dynamics \cite{10.1063/5.0301775}.

Despite the efficiency of the multiset ansatz, two important sources of computational overhead remain.
First, typical multiset calculations employ fixed bond dimensions across all sets that must be chosen sufficiently large to capture the most highly entangled regions of the dynamics.
Since the entanglement structure generally varies strongly between different electronic configurations and throughout the propagation, this choice often results in a substantial fraction of the tensor network carrying larger bond dimension than is required.
Second, in many excitonic problems the population of the multiset wavefunction is concentrated within only a small subset of electronic configurations at any given time.
However, conventional multiset propagators evolve every electronic configuration with equal fidelity, even those that have negligible contributions to the overall dynamics.
These inefficiencies become more pronounced as the system size increases.
For spatially extended systems the number of electronic configurations grows with the size of the lattice while the physically active region (the extent of the exciton wavepacket) at a given time point does not.
Consequently, the fraction of electronic states carrying negligible weight, and therefore the fraction of the computational effort that is wasted, grows with system size. 
As such, a conventional fixed bond dimension propagation becomes progressively less efficient precisely in the regime where large-scale simulation techniques are most needed.

A range of adaptive bond dimension schemes have been developed for single set tensor networks.
In the matrix product state (MPS) and tree tensor network (TTN) communities these include two site time-dependent variational principle based strategies, subspace expansion and Krylov basis schemes, and controlled bond expansions \cite{Bauernfeind2020, PhysRevB.104.214302, doi:10.1021/jacsau.1c00474, PhysRevB.97.045145, Ceruti2022, arxiv.2201.10291, PhysRevLett.133.026401, PhysRevB.94.165116, PhysRevB.102.094315}, while in the multilayer multiconfiguration time-dependent Hartree (ML-MCTDH) community, spawning schemes for single particle functions have been developed. \cite{MendiveTapia2020, PhysRevLett.124.150601, 10.1063/5.0301775}
These approaches provide an efficient mechanism for enlarging the variational manifold where the evolving wavefunction requires it.
However, a corresponding adaptive framework for multiset tensor networks has, to our knowledge, not previously been formulated.

In this manuscript, we present an adaptive multiset TTN (AMTTN) approach that addresses both sources of inefficiency.
First, we derive a local energy-variance spawning criterion for multiset wavefunctions, to arrive at an adaptive bond dimension scheme that naturally accounts for both diagonal and off-diagonal electronic couplings.
Second, we introduce a strategy for identifying weakly populated configurations over the subsequent timestep and temporarily exclude them from explicit propagation while retaining their contribution to the coupled multiset Hamiltonian.
Together, these mechanisms allow computational resources to be dedicated to the dynamically important regions of both the tensor network and electronic state spaces.

Applications of this approach to the excitonic systems illustrate significant reductions in runtime while preserving the accuracy of the dynamics.
When applied to the one dimensional Holstein model, AMTTN provides tightly converged dynamics at a cost one to two orders of magnitude lower than that of published fixed bond dimension multiset MPS calculations\cite{PhysRevLett.123.126601}, while additionally using larger maximum bond dimension for the most challenging regions of the wavefunction. 

At the accuracy required to converge long time exciton dynamics, this approach compares favourably with recently reported GPU accelerated sparse wavefunction calculations of the same model\cite{10.1063/5.0325590} (Appendix~\ref{app:paces}).
For a model of a P3HT:PCBM heterojunction, this approach allows for simulation of chains of 26 and 52 oligothiophene units, two and four times larger than previous tensor network treatments of the system. This allows us to overcome limitations of previous 13 unit simulations\cite{doi:10.1021/acs.jctc.4c00751,10.1063/5.0301775,10.1021/acs.jctc.5c01518}, which fail to accurately capture populations of bound charge separated states as a result of finite size effects.

\section{Theory \label{sec:features}}
\subsection{Multiset Tree Tensor Network Ansatz}

\begin{figure}[t]
\centering
\includegraphics[width=0.5\textwidth]{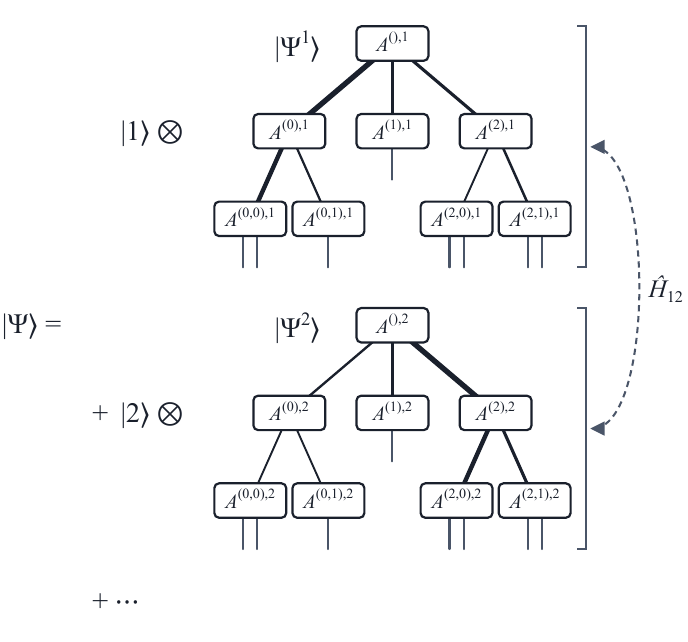}
\vspace{-2em}
\caption{\label{fig:multiset_ansatz}
The multiset tree tensor network ansatz.  The wavefunction is expressed as a sum of direct products of system states $\ket{\alpha}$ and system state dependent tensor networks $\ket{\Psi^{\alpha}}$. Hamiltonian terms coupling different set variables lead to coupling between the tensor networks. Line widths indicate set dependent bond dimensions }
\end{figure}

Throughout this work we consider the multiset tree tensor network (TTN) ansatz \cite{BECK20001, Meyer2003, 10.1002/9783527627400, VENDRELL2011}
\begin{equation} \ket{\Psi(t)} = \sum_{\alpha=1}^{N_{\rm set}} \ket{\alpha} \ket{\Psi^{\alpha}(t)},
\end{equation}
where $\{\ket{\alpha}\}$ denotes a set of $N_{\rm set}$ orthonormal system states and $\ket{\Psi^\alpha(t)}$ represents the environmental wavefunction associated with state $\alpha$.  Here we will use the same notation used in Ref.~\citenum{10.1063/5.0301775}, adapted to the multiset case. Within the multiset representation, each environmental wavefunction $\ket{\Psi^\alpha(t)}$ is represented by an independent TTN with, in order to ensure efficient updating schemes, the same topology, \cite{BECK20001}
\begin{equation}
\ket{\Psi^\alpha(t)} = \sum_{I^{z},i} A^{z,\alpha}_{I^{z}i}(t) \ket{\Psi^{z,\alpha}_{i}(t)} \otimes \ket{\Phi^{z,\alpha}_{I^{z}}(t)},\label{eq:state}
\end{equation}
where $A^{z,\alpha}_{I^{z}i}(t)$ is the coefficient tensor associated with system state $\alpha$ and node $z$ in the tensor network, $\ket{\Phi^{z,\alpha}_{I^{z}}(t)}$ are the corresponding direct product basis functions, the single particle functions (SPFs) associated with the children of node $z$, and $\ket{\Psi^{z,\alpha}_{i}(t)}$ are the corresponding basis functions associated with the remainder of the tree, the single hole functions (SHFs).  For each node $z$ we have a different set of coefficient tensors, single particle functions and single hole functions, that each represent the same overall wavefunction.
Explicitly the direct product basis function at a node $z$, can be written as 
\begin{equation}
    \ket{\Phi^{z,\alpha}_{I^{z}}(t)} = \otimes_k \ket{\phi_{i_k}^{(z,k),\alpha}(t)}
\end{equation}
where $\ket{\phi_{i_k}^{(z,k),\alpha}(t)}$ are the SPFs associated with the $k$-th child node of $z$, namely $(z,k)$, and $I^Z = \{i_1, i_2, \dots\}$ is a multi-index indexing the individual SPFs.  An illustration of the tensor network ansatz is given in Fig.~\ref{fig:multiset_ansatz}.
In the language more common in the MPS literature, the SPFs of node $z$ are isometric basis states associated with subtrees below $z$, the SHFs are the orthonormal environment basis of the remainder of the tree, and Eq.~\ref{eq:state} is the mixed canonical form of the TTN with orthogonality centre at node $z$.

These SPFs and SHFs can be expanded in terms of SPF and SHF functions associated with other nodes on the tensor network. Within the multiset representation, all components share a common tree topology but can in general have independent coefficient tensors, SPFs, SHFs and bond dimensions. 
Detailed overviews of the terminology, single and multiset ans\"atze, and evolution schemes can be found in Refs.~\citenum{WANG2003,MANTHE2008,10.1002/9783527627400,VENDRELL2011, doi:10.1080/00268976.2024.2306881, lindoy_mctdh_1, lindoy_mctdh_2}. 

The multiset TTN ansatz has been successfully applied to problems in nonadiabatic molecular dynamics~\cite{RAAB1999}, exciton dynamics~\cite{PhysRevLett.123.126601,10.1063/1.5051498,10.1063/5.0301775}, and related tensor network approaches to dissipative quantum dynamics~\cite{PhysRevLett.123.100502,10.1063/5.0266136}.
A central advantage of the ansatz is that each electronic configuration has an independent representation of the environment degrees of freedom and therefore an independent entanglement structure.
As a consequence, the optimal bond dimensions can differ substantially between multiset components and may vary strongly throughout the time evolution.
Furthermore, many applications exhibit highly non-uniform populations across electronic configurations, such that only a subset of multiset components make a significant contribution to the dynamics at any given time.

\subsection{Adaptive Multiset Control}
\subsubsection{Energy Variance Based Spawning of Multiset Wavefunctions}
We extend the energy variance formalism introduced in Ref.~\citenum{MendiveTapia2020} and subsequently formulated in the mixed canonical gauge~\cite{Lubich2014a,Kloss2017a, Bauernfeind2020,Kloss2020,Ceruti2021, lindoy_thesis, lindoy_mctdh_1, lindoy_mctdh_2} in Ref.~\citenum{10.1063/5.0301775} to the multiset TTN ansatz of Eqs.~1 and 2.  
Within the multiset formalism the dynamics is generated by a Hamiltonian of the form 
\begin{equation}
\hat H = \sum_{\alpha,\beta} \ket{\alpha}\bra{\beta} \otimes \hat H_{\alpha\beta},\label{eq:H1}
\end{equation}
where the operators $\hat H_{\alpha\beta}$ act on the common set of environmental degrees of freedom.  For efficient implementation each block of the multiset Hamiltonian is taken in sum of product (SOP) form,
\begin{equation}
\hat H_{\alpha\beta} = \sum_{r} c_r^{\alpha\beta} \bigotimes_{\mu=1}^{D} \hat h_{\mu,r}^{\alpha\beta},
\label{eq:ms-sop}
\end{equation}
where $\mu$ labels the $D$ physical degrees of freedom and each $\hat h_{\mu,r}^{\alpha\beta}$ acts on one of them.

At a given node $z$ of the tree, every product term factorises into a part acting on the physical degrees of freedom below $z$ and a part acting on the remainder of the tree,
\begin{equation}
    \hat H_{z,\alpha\beta} = \sum_r c_r^{z,\alpha\beta}\, \hat h_r^{z,\alpha\beta}\, \hat H_r^{z,\alpha\beta},
\label{eq:ms-spo-mfo}
\end{equation}
with $\hat h_r^{z,\alpha\beta}$ the single-particle operators (SPO) and $\hat H_r^{z,\alpha\beta}$ the mean-field operators (MFO).  Both types of operators are direct products of the primitive operators (Eq.~(\ref{eq:ms-sop})) and are built using an analogous recurrence relation to the single set case.  The SPOs of a node are the product of SPOs of its children,
\begin{equation}
\hat h_r^{z,\alpha\beta} = \bigotimes_k \hat h_r^{(z,k),\alpha\beta},
\label{eq:ms-spo-rec}
\end{equation}
while the MFOs of a child are obtained from the MFOs of its parent together with the SPOs of its siblings,
\begin{equation}
\hat H_r^{(z,k),\alpha\beta} = \hat H_r^{z,\alpha\beta} \bigotimes_{j\neq k} \hat h_r^{(z,j),\alpha\beta}.
\label{eq:ms-mfo-rec}
\end{equation}

Projecting the SPO and MFO operators onto the local bases associated with node $z$ of the tree gives the corresponding matrices:
\begin{align}
\left[\boldsymbol{h}_r^{z,\alpha\beta}(t)\right]_{IJ} &= \bra{\Phi_I^{z,\alpha}(t)}\, \hat h_r^{z,\alpha\beta}\, |\Phi_J^{z,\beta}(t)\rangle, \label{eq:ms-spo-mat}\\
\left[\boldsymbol{H}_r^{z,\alpha\beta}(t)\right]_{ij} &= \langle\Psi_i^{z,\alpha}(t)|\, \hat H_r^{z,\alpha\beta}\, |\Psi_j^{z,\beta}(t)\rangle. \label{eq:ms-mfo-mat}
\end{align}

As in the single set formulation, the adaptive expansion procedure is derived starting from a local-in-time Dirac-Frenkel error metric\cite{MendiveTapia2020, PhysRevLett.124.150601, 10.1063/5.0301775}.
To do this  we consider two wavefunctions $\ket{\Psi_e(t)} = \sum_\alpha \ket{\alpha} \ket{{\Psi}_{e}^{\alpha}(t)}$ and $\ket{\Psi_{M}(t)}= \sum_\alpha \ket{\alpha} \ket{{\Psi}_{M}^{\alpha}(t)}$ corresponding to the exact solution of the Schr\"odinger equation, and the approximate TTN solution obtained with a given set of bond dimensions throughout the tensor. 
Due to the orthogonality of the system basis states, $\braket{\alpha}{\beta}=\delta_{\alpha\beta}$, the global error between these two states decomposes into a sum over independent contributions from each multiset component.  
\begin{equation}
E =\left\| \ket{\dot{\Psi}_{e}(t)} - \ket{\dot{\Psi}_{M}(t)} \right\|^2 =\sum_\alpha E^\alpha,
\end{equation}
where 
\begin{equation}
E^\alpha = \left\| \ket{\dot{\Psi}_{e}^{\alpha}(t)} - \ket{\dot{\Psi}_{M}^{\alpha}(t)} \right\|^2 .
\end{equation}
This allows for the spawning problem to be formulated independently for each multiset tensor network, providing a means to adapt the bond dimensions associated with different system configurations independently.

Following Refs.~\citenum{MendiveTapia2020,10.1063/5.0301775}, we consider how to optimally extend the SPF basis associated with node $(z,k)$ of the TTN corresponding to multiset component $\alpha$, by introducing a new set of single-particle functions $\ket{\tilde{\phi}_i^{(z, k), \alpha}}$ associated with node $(z,k)$ on set $\alpha$ that are mutually orthogonal to the existing single-particle functions at that set and node. 
Introducing these additional unoccupied SPFs leaves the represented wavefunction unchanged, but enlarges the tangent space available for subsequent evolution.
Repeating the analysis of the single set case \cite{10.1063/5.0301775} then yields an energy error 
\begin{equation}
\Delta_d^{(z,k),\alpha} = E^\alpha-\tilde E^\alpha = \sum_m \bra{\tilde\phi_m^{(z,k),\alpha}} \hat\Delta^{(z,k),\alpha} \ket{\tilde\phi_m^{(z,k),\alpha}},
\end{equation}
where $\tilde E^\alpha =\left\| \ket{\dot{\Psi}^\alpha_{e}(t)} - \ket{\dot{\tilde{\Psi}}^\alpha_{M}(t)} \right\|^2$, is the error associated with set $\alpha$ observed after augmenting the wavefunction with the new set of single-particle functions. In this expression we have introduced the operator
\begin{widetext}
\begin{equation}
\begin{split}
\hat\Delta^{(z,k),\alpha} = (1-P^{(z,k),\alpha}) \sum_{i,I^{z\setminus k}} \braket{\Theta_{iI^{z\setminus k}}^{(z,k),\alpha}} {\dot\Psi_e^\alpha-\dot\Psi_M^\alpha} 
\braket{\dot\Psi_e^\alpha-\dot\Psi_M^\alpha} {\Theta_{iI^{z\setminus k}}^{(z,k),\alpha}} (1-P^{(z,k),\alpha}),
\end{split}
\end{equation}
\end{widetext}
where
\begin{equation} 
\ket{\Theta_{iI^{z\setminus k}}^{(z,k),\alpha}} = \ket{\Psi_i^{z,\alpha}} \otimes \ket{\Phi_{I^{z\setminus k}}^{z\setminus k,\alpha}} 
\end{equation}
denotes the orthonormal basis for all degrees of freedom of TTN $\alpha$ that are not descendants of node $(z,k)$, and $z\setminus k$ indicates that we are constructing the direct product SPFs associated with all children of node $z$ except for node $k$
\begin{equation}
    \ket{\Phi_{I^{z\setminus k}}^{z\setminus k,\alpha}} = \bigotimes_{j\neq k}  \ket{\phi_{i_j}^{(z,j),\alpha}(t)}
\end{equation}
and the projector 
\begin{equation}
P^{(z,k),\alpha} = \mat A^{(z,k),\alpha}\bigl(\mat A^{(z,k),\alpha}\bigr)^\dagger,
\label{eq:ms-spf-projector}
\end{equation}
is obtained from a matricisation of the site tensor $A^{(z,k),\alpha}$, where the uncontracted index is the index associated with the bond we are attempting to expand.

For a given system state, $\alpha$, the exact time evolution is governed by projecting the Schr\"odinger equation onto the electronic state $\bra{\alpha}$ which gives
\begin{equation}
\ket{\dot\Psi_e^{\alpha}(t)} = -i \left( \hat H_{\alpha\alpha}\ket{\Psi_e^{\alpha}(t)} + \sum_{\beta\neq\alpha} \hat H_{\alpha\beta} \ket{\Psi_e^{\beta}(t)} \right), \label{eq:ms-eom}
\end{equation}
(where we use units in which $\hbar=1$) containing contributions associated with all coupled system states.

 Within the projector splitting integrator approach, the time derivative of the state remains entirely within the tangent space associated with the current SPF basis\cite{Lubich2014a,Kloss2017a}, that is, 
 \begin{equation} 
 (1-P^{(z,k),\alpha}) \braket{\Theta_{iI^{z\setminus k}}^{(z,k),\alpha}}{\dot\Psi_M^\alpha(t_0)} = 0.
 \end{equation} 
 Combining this observation, enforcing that $\ket{\Psi_e^\beta(t_0)}=\ket{\Psi_M^\beta(t_0)}$ at $t_0$ for all $\beta$, and Eq.~(\ref{eq:ms-eom}) gives
\begin{equation}
\begin{split}
(1-&P^{(z,k),\alpha}) \braket{\Theta_{iI^{z\setminus k}}^{(z,k),\alpha}}{\dot\Psi_e^\alpha \!-\! \dot\Psi_M^\alpha}  \\ = &\!-i (1\!-\!P^{(z,k),\alpha}) \sum_{\beta} \bra{\Theta_{iI^{z\setminus k}}^{(z,k),\alpha}}\hat H_{\alpha\beta} \ket{\Psi_M^\beta} .
\end{split}
\end{equation} 
Substituting this expression into the variance operator gives 
\begin{widetext}
\begin{equation}
\begin{split}
\hat\Delta^{(z,k),\alpha} = &(1-P^{(z,k),\alpha}) \sum_{i,I^{z\setminus k}} \braket{\Theta_{iI^{z\setminus k}}^{(z,k),\alpha}} {\sum_\beta \hat H_{\alpha\beta}  \Psi_M^\beta} \braket{\sum_\beta \hat H_{\alpha\beta} \Psi_M^\beta} {\Theta_{iI^{z\setminus k}}^{(z,k),\alpha}} (1-P^{(z,k),\alpha}) \\
 = & (1-P^{(z,k),\alpha}) M^{(z,k), \alpha \dagger}M^{(z,k), \alpha}(1-P^{(z,k),\alpha}).
\end{split}
\end{equation}
\end{widetext}
with 
\begin{equation}
    \left[M^{(z,k), \alpha}\right]_{(i,I^{z\setminus k}),m} =  \big \langle m^{(z,k),\alpha}\big|\big\langle\Theta_{iI^{z\setminus k}}^{(z,k),\alpha} \big| \sum_\beta \hat H_{\alpha\beta}  \Psi_M^\beta\big\rangle,
\end{equation}
where $ \{ \ket{m^{(z,k),\alpha}}\}$ denotes an orthonormal basis for the Hilbert space associated with node $(z,k)$.

This operator has the same structure as the single set TTN operator defined in Ref.~\citenum{10.1063/5.0301775}, however, with coupling between set variables through the Hamiltonian terms, $\sum_\beta \hat H_{\alpha\beta}\ket{\Psi_M^\beta}$.
As the off-diagonal blocks $\hat H_{\alpha\beta}$ enter the variance of component $\alpha$ weighted by the coupled wavefunctions $\ket{\Psi_M^\beta}$, a configuration that is unpopulated but adjacent to a populated one acquires non-zero values of $\hat\Delta^{(z,k),\alpha}$ before population reaches it.  This enables expansion of bond dimension ahead of the wavepacket arriving at this set variable rather than in response to it. 

The dominant eigenvectors of $\hat\Delta^{(z,k),\alpha}$ are the optimal spawning directions for node $(z,k)$\cite{MendiveTapia2020, 10.1063/5.0301775}.
However, explicitly constructing the variance operator requires forming a two site object, reintroducing the computational bottleneck that we aim to avoid using the subspace expansion approach, as in the two site integrators.
Instead, as in Ref.~\citenum{10.1063/5.0301775}, we exploit the structure of the Hamiltonian to obtain a matrix free representation of the variance operator and determine its dominant eigenpairs iteratively.

Using that each operator $\hat H_{\alpha\beta}$ is expressed as a sum of product terms, we can expand the action of the multiset Hamiltonian on each state vector.  
Doing so we obtain an expression for the matrices $M^{(z,k),\alpha}$, 
\begin{widetext}
\begin{equation}
\left[M^{(z,k),\alpha}\right]_{(i,I^{z\setminus k}),m}
\!=\!\sum_{\beta}\sum_{r} c_r^{\alpha\beta}\!\!\!\!\!
\sum_{j,\,J^{z\setminus k},\,m'}\!\!
\!\!\!\!\left[\boldsymbol{H}_r^{z,\alpha\beta}\right]_{ij}\,
\!\left[\boldsymbol{h}_r^{z\setminus k,\alpha\beta}\right]_{I^{z\setminus k}J^{z\setminus k}}\,
\!\left[\boldsymbol{h}_r^{(z,k),\alpha\beta}\right]_{mm'}\,
\!\!A^{z,\beta}_{(J^{z\setminus k},m')\,j},
\label{eq:M-explicit}
\end{equation}
\end{widetext}
where $\boldsymbol{h}_r^{z\setminus k,\alpha\beta}=\bigotimes_{j\neq k}\boldsymbol{h}_r^{(z,j),\alpha\beta}$ is a direct product of SPO matrices of the siblings of $(z,k)$, and $\boldsymbol{H}_r^{z,\alpha\beta}$, $\boldsymbol{h}_r^{(z,k),\alpha\beta}$ are the MFO and SPO matrices(Eqs.~\ref{eq:ms-spo-mat} and \ref{eq:ms-mfo-mat}). From this, we can evaluate the action of these matrices on a general vector, $v$, as
\begin{align}
M^{(z,k),\alpha} v &= \sum_{\beta,r} c_r^{\alpha\beta}\,
\bigl(\boldsymbol{H}_r^{z,\alpha\beta}\otimes\boldsymbol{h}_r^{z\setminus k,\alpha\beta}\bigr)\,
A^{z,\beta}\,\bigl[(\boldsymbol{h}_r^{(z,k),\alpha\beta})^{T} v\bigr],
\label{eq:M-apply}
\end{align}
where each term involves contractions that have the same asymptotic cost as the cost of terms in the effective Hamiltonian application. Analogous expressions can be obtained for evaluating the action of $M^{(z,k),\alpha\dagger}$. 
The presence of an efficient approach for implementing the matrix-vector products allows for the extraction of dominant spawning directions using Krylov subspace techniques\cite{Saad2011}.  
In all calculations performed in this manuscript we make use of an Arnoldi iteration to extract a small number of the dominant eigenpairs in order to spawn SPFs and SHFs.
From these eigenpairs, we can construct the singular values of $M^{(z,k),\alpha}(1-P^{(z,k),\alpha})$, $\sigma^{(z,k),\alpha}$ and we spawn new SPFs and SHFs whenever
\begin{equation}
    \sigma^{(z,k),\alpha} \delta t \geq \rm tol,\label{eq:spawning_criterion}
\end{equation}
where $\delta t$ is the timestep used throughout the step, and ${\rm tol}$ is a convergence parameter representing the singular value spawning tolerance.  The singular values, $\sigma^{(z,k),\alpha}$, represent the magnitude of the wavefunction derivative along this proposed direction (outside of the current local basis) and consequently $\sigma^{(z,k),\alpha} \delta t$ estimates the first order amplitude that would be generated along this direction in a timestep. 
Using this criterion we allow the bond dimension, $\chi$, associated with each bond in the multiset tensor network to expand up to a maximum bond dimension, $\chi_{\rm max}$.

\subsubsection{Adaptive Identification of Inactive Sets \label{sec:set_freezing}}
For  systems that contain a large number of coupled electronic configurations, the population of the wavefunction is often concentrated in a small subset of electronic configurations.
Applying a fixed bond dimension multiset ansatz in such cases can lead to the cost of propagation being dominated by electronic configurations that provide negligible weight to the wavefunction.
Motivated by other approaches that adaptively determine the active state space in wavepacket dynamics \cite{10.1063/5.0325590, PhysRevB.101.174315,PETTEY2006443,10.1063/1.1560636,PhysRevLett.110.263202,10.1021/acs.jctc.7b00021, 10.1021/jp994174i}, here we outline a power series expansion based approach for dynamically determining the relevant set variables to propagate within each time step.  This strategy aims to exploit the finite propagation speed of the exciton dynamics, over a single timestep population can only reach configurations that are connected to the currently populated region through a small number of Hamiltonian applications.

The instantaneous population of each set variable
\begin{equation}
    w_\alpha(t) = \braket{\Psi^\alpha(t)} {\Psi^\alpha(t)},
\end{equation}
is, on its own, insufficient to determine which sets are relevant through the next step of the dynamical evolution.  
In order to obtain a more reliable estimate of dynamically relevant states, we construct a short-time prediction of the evolution of each set variable using the effective Hamiltonian available at the current orthogonality centre of the multiset TTN. 

During the projector splitting integrator propagation~\cite{Lubich2014a,Kloss2017a, Bauernfeind2020,Kloss2020,Ceruti2021,lindoy_mctdh_2}, repeated applications of the effective Hamiltonian 
\begin{equation}
[\hat H_{\rm eff}^{z,\alpha\beta}]_{(I^zi,J^zj)} =  \bra{\Phi_{I^z}^{z,\alpha}\otimes\Psi^{z,\alpha}_i}\hat H_{\alpha\beta} \ket{\Phi_{J^z}^{z,\beta}\otimes\Psi^{z,\beta}_j}
\end{equation}
to the local coefficient vector $\ket{\psi^z(t)}$, which has elements $\psi_{I^zi}^{z,\alpha}(t) = \braket{\Phi_{I^z}^{z,\alpha}\otimes\Psi^{z,\alpha}_i}{\Psi^\alpha(t)}$, are evaluated as part of the evolution of the orthogonality centre. 
Reusing this information, we form a truncated power series approximation to estimate the time evolved coefficient vector, 
\begin{equation} 
\ket{\tilde \psi(t+\delta t)} = \sum_{j=0}^{K} \frac{(-i\delta t)^j}{j!} \hat H_{\rm eff}^{\,j}{\psi^z(t)},
\end{equation} 
where $\delta t$ is the time step (here taken to be the full timestep associated with the forward and backward sweep steps of the projector splitting integrator) and $K$ is the order of expansion.
From this approximately propagated state, we obtain an approximate population of
\begin{equation} 
\tilde w_\alpha(t+\delta t) = \braket{\tilde \psi^z_\alpha(t+\delta t)}{\tilde \psi^z_\alpha(t+\delta t)}, 
\end{equation} 
that accounts for population transfer from occupied to unoccupied states.

A set variable is classified as dynamically active if either its instantaneous population or its predicted population exceeds the corresponding tolerance, 
\begin{equation} 
w_\alpha(t) \geq {\rm tol}_{w} \qquad \text{or} \qquad \tilde{w}_\alpha(t+\delta t) \geq {\rm tol}_{\tilde w}. \label{eq:set_activity_criterion} 
\end{equation} 
Only active set variables are propagated during the subsequent time step. 
Inactive variables are kept fixed until being reconsidered at the next timestep. 

The criterion in Eq.~\ref{eq:set_activity_criterion} is distinct from the spawning criterion in Eq.~\ref{eq:spawning_criterion}, and both can be applied in parallel. 
The spawning criterion controls expansion of the local SPF and SHF bases within the multiset tensor network. 
In contrast,  Eq.~\ref{eq:set_activity_criterion} determines which set variables are propagated during a timestep. 
These two tolerances control two distinct approximations that can be converged independently.
We refer to the multiset TTN with adaptive spawning, and optionally set freezing, as the adaptive multiset tree tensor network (AMTTN) approach.
\section{Results}

\begin{figure}[t]
\centering
\includegraphics[width=0.5\textwidth]{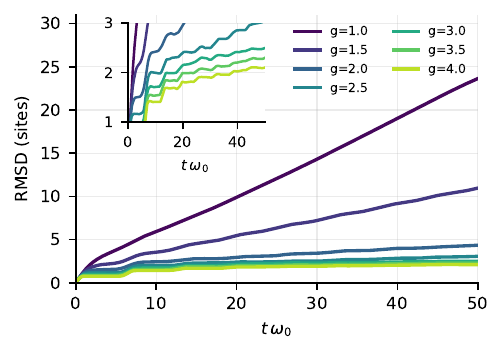}
\vspace{-2em}
\caption{\label{fig:holstein_rmsd}
The exciton root mean square displacement against time for various couplings $g$ of the 1D Holstein model, computed using AMTTN.  The inset shows the dynamics for the strong-coupling parameter regime.}
\end{figure}

\subsection{Exciton Dynamics in the One Dimensional Holstein Model}
We first consider the one dimensional Holstein model.  
The real time dynamics of this model was previously studied using multiset matrix product state methods by Kloss \emph{et al.} ~\citenum{PhysRevLett.123.126601}.  The Holstein model\cite{HOLSTEIN1959325} describes the coupling of  mobile electronic excitations to local vibrational degrees of freedom. In the site basis, the Hamiltonian is given by
\begin{equation}
\begin{split}
\hat H =& -J \sum_{\langle i,j\rangle} \left( \hat a_i^\dagger \hat a_j + \hat a_j^\dagger \hat a_i \right) + \omega_0 \sum_i \hat b_i^\dagger \hat b_i \\& + g\omega_0 \sum_i \hat a_i^\dagger \hat a_i \left( \hat b_i^\dagger + \hat b_i \right),
\end{split}
\end{equation}
where $\hat a_i^\dagger$ ($\hat a_i$) is the creation (annihilation) operator for an exciton on lattice site $i$, $\hat b_i^\dagger$ ($\hat b_i$) is the creation (annihilation) operator for the local vibrational mode of frequency $\omega_0$, $J$ is the nearest neighbour electronic hopping term, and $g$ denotes the dimensionless exciton-phonon coupling strength.

All Holstein model calculations that are presented below use open boundary conditions, $J=\omega_0$, a single electronic excitation initialised on the central site with all vibrational modes in their ground state, a time step of $\mathrm{d}t=0.05\,\omega_0^{-1}$ and propagation to $t\omega_0=50$ in order to allow for comparison with prior studies\cite{PhysRevLett.123.126601, 10.1063/5.0325590}.  
Local boson spaces are truncated using the same parameters as given in Ref.~\citenum{PhysRevLett.123.126601}, that is we use 16 states for $g=1$ and $1.5$, 64 states for $g=2$, $2.5$ and $3$, and 128 states for $g=3.5$ and $4$.
Likewise we use the lattice sizes, $N$, from Ref.~\citenum{PhysRevLett.123.126601}, except for the system size scaling results which vary $N$ over the range $10$ to $250$.
Throughout this section, the dynamics are characterised by the exciton root-mean-square displacement,
\begin{equation}
\mathrm{RMSD}(t)= \sqrt{ \sum_i (i-i_0)^2 \langle \hat a_i^\dagger(t) \hat a_i(t) \rangle},
\end{equation}
where $i_0$ denotes the initially excited site.

Figure~\ref{fig:holstein_rmsd} presents $\mathrm{RMSD}(t)$ for the full range of couplings considered obtained using AMTTN with maximum allowed bond dimensions, $\chi=128$.  The results obtained here agree well with those of Ref.~\citenum{PhysRevLett.123.126601}.  At weak coupling $\mathrm{RMSD}(t)$ exhibits essentially linear growth with time, indicating near ballistic transport for the weakly dressed exciton.  As the coupling increases the growth is suppressed, indicating the formation of polarons with larger effective masses \cite{PhysRevLett.123.126601}. 

\begin{figure}[t]
\centering
\includegraphics[width=0.5\textwidth]{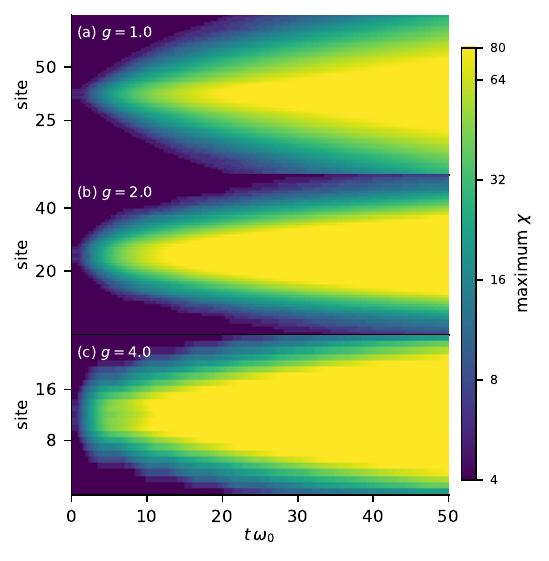}
\vspace{-2em}
\caption{\label{fig:holstein_bonddim}
Maximum bond dimension for each set index as a function of time, for $\chi=80$ and spawning tolerance of $10^{-4}$ at (a) $g=1$, (b) $g=2$, and (c) $g=4$.}
\end{figure}

The motivation for the adaptive bond dimension control is illustrated  in Fig.~\ref{fig:holstein_bonddim}. This figure shows the evolution of the maximum bond dimension associated with each multiset component obtained from simulations with $\chi_{\max}=80$, spawning tolerance $10^{-4}$, and illustrates several key points.
First, the bond dimensions required are strongly heterogeneous across set variables, with only a subset of components requiring substantial bond dimension at any given time.
Second, this growth occurs dynamically as the excitation propagates through the lattice.
Finally, the degree of heterogeneity increases with coupling strength, where strong coupling to local vibrations suppresses exciton transport and leads to a small number of highly correlated components while set variables corresponding to electronic states far away see little to no bond dimension growth.

\begin{figure}[t]
\centering
\includegraphics[width=0.475\textwidth]{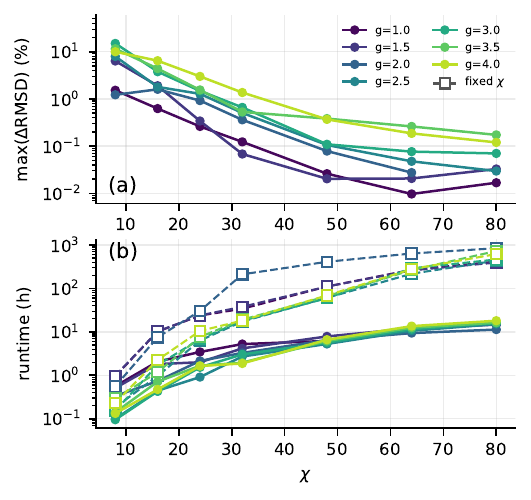}
\vspace{-2em}
\caption{\label{fig:holstein_chi}
(a)~Maximum RMSD error against the $\chi=128$, spawning tolerance $10^{-6}$ reference simulation for each coupling as a function of the maximum $\chi$ used for the simulation.  (b)~Total single core runtime to reach $t\omega_0=50$ for the same runs (circles) and for the equivalent fixed-$\chi$ simulation.  
Due to the large runtimes associated with the large fixed $\chi$ simulations, the fixed $\chi$ runtimes are estimated by extrapolating the average, near constant per step cost of short-time ($t < 5$) fixed $\chi$ runs. 
All (non reference) adaptive simulations were run with a spawning tolerance of $10^{-4}$, and without the use of set freezing. }
\end{figure}

In Fig.~\ref{fig:holstein_chi} we consider the convergence and runtime behaviour of AMTTN. 
All CPU timings reported in this work are single core wall times on an Intel Xeon Gold 6338 processor.
Note that all jobs were run on shared nodes and therefore may exhibit variation due to node load.
Panel a) presents the maximum RMSD error relative to AMTTN calculations obtained with $\chi=128$ and spawning tolerance $\mathrm{tol}=10^{-6}$ calculations for a range of maximum allowed bond dimensions.  
For all coupling strengths, the error decreases systematically as the maximum $\chi$ increases, indicating convergence to the reference.   

The corresponding runtimes are shown in Panel (b) of Fig.~\ref{fig:holstein_chi}.
For a given maximum $\chi$ the adaptive calculations are consistently faster than fixed $\chi$ simulations, with speedups that grow with $\chi$ and reach one to two orders of magnitude for $\chi\gtrsim24$.
The largest gains occur at large bond dimensions, where the cost of full fixed $\chi$ evaluation is largest, and in regimes where only a small fraction of the multiset components require significant bond dimension growth, e.g. small $g$, consistent with the heterogeneous bond dimension growth seen in Fig.~\ref{fig:holstein_bonddim}. 
A more detailed comparison of the convergence (with respect to $\chi$ and spawning tolerances) and runtime of the adaptive multiset approach is given in Appendix~\ref{appendix:cpu_runtime} for $g=1$, $2$, and $4$, together with a set of simulation parameters that are suitable for obtaining sub percent convergence of the RMSD (well within the thickness of the lines of Fig.~\ref{fig:holstein_rmsd}) across all considered $g$.
In Appendix~\ref{app:paces}, we additionally compare AMTTN with the GPU-accelerated sparse wavefunction method PACES\cite{10.1063/5.0325590} for the strongly coupled $g=4$ chain.

We next consider how the computational cost of AMTTN varies with the number of sites in the Holstein model, and additionally assess the benefits of the set freezing strategy introduced in Sec.~\ref{sec:set_freezing}.
For set freezing simulations, we use $K=4$, which allows for any set that is connected to the current populated set of configurations by $K=4$ Hamiltonian applications to be active in any given timestep.
Additionally, we find well converged results, corresponding to relative RMSD errors of $< 0.01\%$, with a population threshold of $10^{-12}$  and a predicted population threshold of $10^{-8}$.
This corresponds to treating all sets with a population of $10^{-12}$ or greater or those connected to the populated sets by at most $K=4$ Hamiltonian applications and a predicted population of $10^{-8}$ as active.

\begin{figure}[t]
\centering
\includegraphics[width=0.475\textwidth]{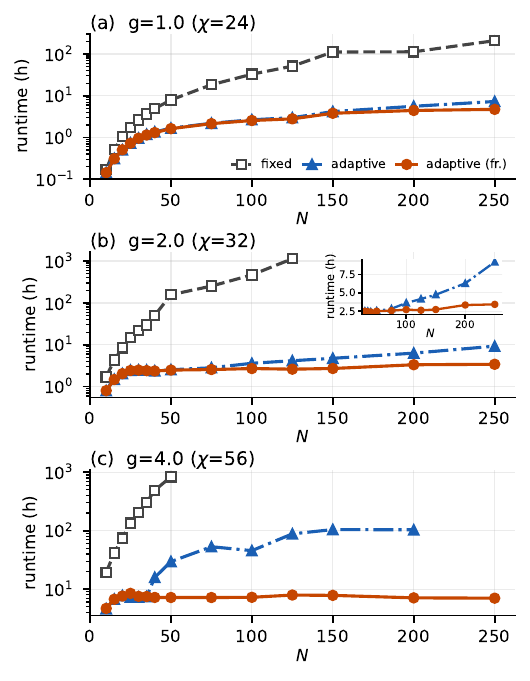}
\vspace{-2em}
\caption{\label{fig:holstein_nscaling}
Total runtime to $t\omega_0=50$ for various Holstein model lengths $N$ for fixed $\chi$, adaptive $\chi$, and adaptive $\chi$ with set freezing
(a)~$g=1.0$ ($\chi_{\max}=24$),
(b)~$g=2.0$ ($\chi_{\max}=32$),
(c)~$g=4.0$ ($\chi_{\max}=56$). The insets show a linear scale comparison of the adaptive simulations using and not using set freezing at large $N$.
Due to the long runtimes involved fixed $\chi$ results for $g=2$ and $g=4$ are extrapolated from the average, near constant runtime per step obtained from shorter $\omega_0 t_{\rm max} \leq 5$ simulation.}
\end{figure}

Fig.~\ref{fig:holstein_nscaling} compares the total simulation time of fixed-$\chi$, adaptive-$\chi$, and adaptive-$\chi$ calculations using set freezing.  Across the system sizes considered the adaptive bond dimension approach substantially reduces runtime compared to the fixed $\chi$ approach.  This reduction arises due to large bond dimensions only arising for multiset components and bonds that require them, rather than being uniformly imposed across the wavefunction.

Set freezing further reduces the runtime by excluding configurations that provide negligible contribution to the evolution.  At small $N$ the overhead associated with identifying inactive sets is larger than any potential benefit from set freezing as the adaptive bond dimension scheme already reduces the cost associated with weakly populated configurations.  
However, the absolute saving from set freezing increases with system size as the local propagation requires the action of a matrix exponential on a coefficient vector whose dimension grows linearly with the number of set variables. For sufficiently large $N$, the cost associated with inactive components can become significant.

To quantify the observed system size dependence we fit the runtime values to a power law, $ \text{runtime}(N) \sim N^{\alpha}$.  For the fixed $\chi$ simulations we find exponents of $\alpha\approx2.0$ at $g=1$ and $\approx2.1$ at $g=2$, this approximately quadratic growth is consistent with expectation due to the linear growth in both the number of multiset components and the number of tensors in each component.
For $g=4$, a fit over $25\leq N\leq50$ gives a larger exponent of $\alpha\approx2.7$. This larger exponent may arise from an increase in the cost of Krylov subspace integration as the dimension and spectral range of the effective generators grow.

The use of adaptive bond dimension control considerably reduces this dependence on system size, giving $\alpha\approx1.0$, $0.8$, and $1.2$ at $g=1$, $2$ and $4$, respectively. This reduction arises as large bond dimensions are required only for a subset of bonds within each multiset component, only those associated with vibrational modes near the corresponding electronic configuration.  
Set freezing further reduces the exponents, giving $\alpha\approx0.7$, $0.2$ and $0.0$ at $g=1$, $2$, and $4$, respectively.
For the systems considered, the combined scheme exhibits runtimes that depend sublinearly on system size, and at strong coupling become largely independent of $N$ for $N\gtrsim30$.
Over the system sizes and propagation times considered, the runtime is controlled primarily by the dynamically active region with only a weak dependence on the full lattice size, in contrast to the approximately quadratic scaling of the fixed bond dimension multiset calculations.  

The coupling dependence of the set freezing results can be attributed to the spatial extent of the excitation in the three cases.  At weak coupling, the excitation propagates nearly ballistically and occupies an increasing number of electronic configurations with time. 
The size of the dynamical region, and therefore cost of evolving the multiset tensor network, grows throughout the simulation and for sufficiently long times may lead to $\mathcal{O}(N)$ active sets.
At strong coupling, exciton phonon interactions localise the excitation within a small subset of configurations.  Additional configurations introduced as $N$ increases remain weakly populated and can be excluded from propagation.  Over the system sizes and propagation times considered the cost of the adaptive scheme with set freezing is largely determined by the size of the dynamically active region with only a weak dependence on the full lattice size.

These results demonstrate the complementarity of the two adaptive strategies. Adaptive bond dimensions reduce the cost of representing each multiset component, while set freezing reduces the cost of propagation and integration overhead associated with inactive components.

AMTTN can also benefit from GPU acceleration.  
For $g=4$ with $\chi_{max}=80$ and a spawning tolerance of $10^{-6}$, a hybrid CPU/GPU calculation on a single A100 GPU is 21 times faster than the same calculation on a single CPU core. Additionally, the adaptive scheme remains faster than fixed bond dimension GPU simulations at the same maximum bond dimension.  Details of the implementation and timings are given in Appendix~\ref{app:gpu_timings}.

\subsection{Charge Separation at a Donor-Acceptor Heterojunction}

\begin{figure}[t]
\includegraphics[width=0.5\textwidth]{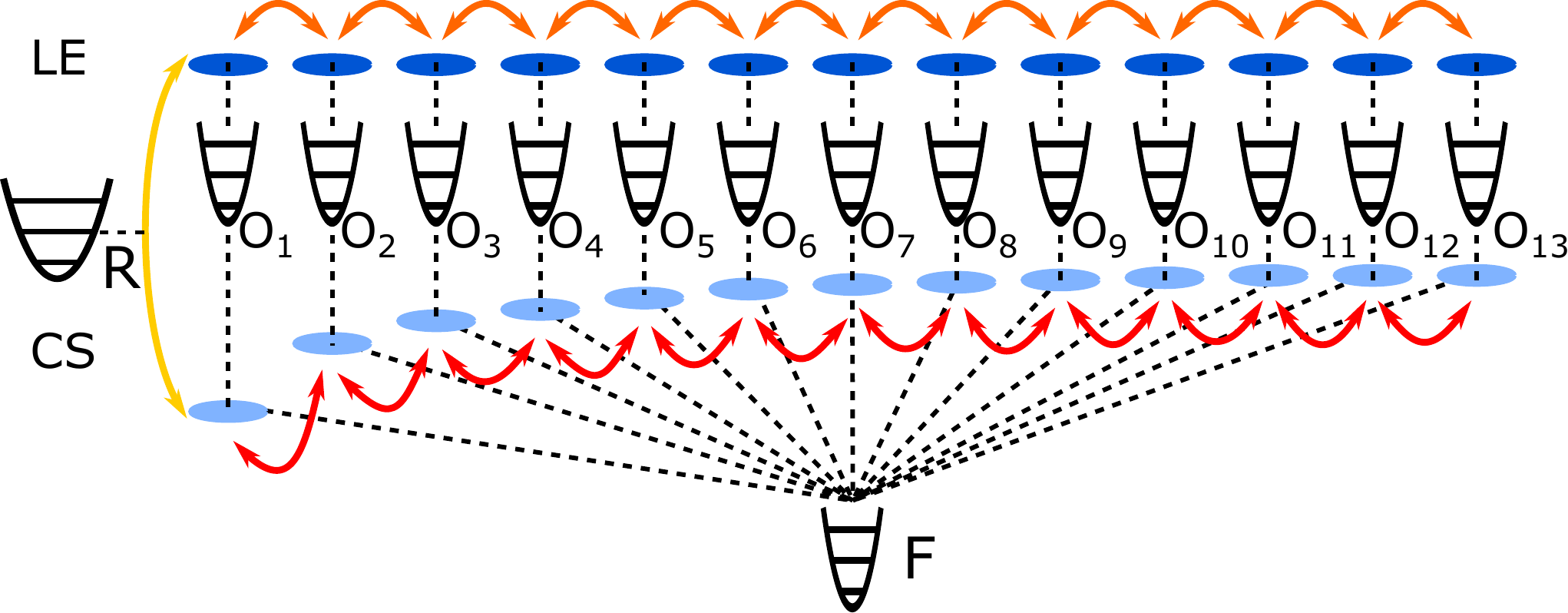} \vspace{-1em}
\caption{ \label{fig:p3ht_pcbm_schematic} Illustration of the interactions present in the P3HT:PCBM heterojunction model, ellipses represent electronic states, parabolas represent vibrational modes, arrows show electronic state couplings, and dotted lines vibronic couplings. The 13 unit model contains 26 total electronic states, 13 local exciton states and 13 charge-separated states\cite{doi:10.1021/acs.jpclett.2c01928,doi:10.1021/ja4093874,doi:10.1021/acs.jctc.4c00751, doi:10.1021/acs.jpclett.5b00336}. For each oligothiophene we consider eight vibrational modes ($O$ modes) that couple to both local exciton and charge separated states associated to this molecule.  Additionally, a set of eight vibrational modes associated with the fullerene acceptor system ($F$ modes) couple to all charge separated states.  Finally, interactions between the charge separated and local exciton states of the oligothiophene at the junction are mediated via molecular vibrations of the $R$ mode.}  
\end{figure}

As a large-scale application, we consider the separation of an exciton into free charge carriers at a donor acceptor interface.  Here we use a linear vibronic coupling model of a $n$-oligothiophene donor-C$_{60}$ fullerene acceptor system (P3HT:PCBM),~\cite{doi:10.1021/acs.jpclett.2c01928,doi:10.1021/ja4093874,doi:10.1021/acs.jctc.4c00751,doi:10.1021/acs.jpclett.5b00336}.
This model has emerged as a challenging system for tensor network approaches due to its large number of coupled electronic and vibrational degrees of freedom, complex nature of the interactions present, and the strongly state dependent vibronic correlations that develop during the charge-separation process \cite{10.1063/5.0301775,10.1021/acs.jctc.5c01518,doi:10.1021/acs.jpclett.5b00336}.

Previous studies have reported noticeable differences between dynamics obtained using MPS and ML-MCTDH based wavefunction ans\"atze~\cite{doi:10.1021/acs.jctc.4c00751}, although several subsequent studies have shown that converged results can be obtained through the use of the multiset ansatz \cite{10.1063/5.0301775} or considerably higher-bond-dimension single set simulations \cite{10.1063/5.0301775, 10.1021/acs.jctc.5c01518}.  These previous tensor network and ML-MCTDH studies have focused on a 13-unit oligothiophene chain, owing to the considerable computational cost associated with the large number of coupled electronic and vibrational degrees of freedom within the problem.  However, the timescales associated with charge transfer dynamics are comparable to the time required for the wavepacket to reach the chain boundary in the charge separated manifold, raising the possibility that finite size effects influence the predicted long time dynamics.  Before considering larger systems, we first validate AMTTN against the established 13-unit benchmark.

\begin{figure}[t]
\centering
\includegraphics[width=0.5\textwidth]{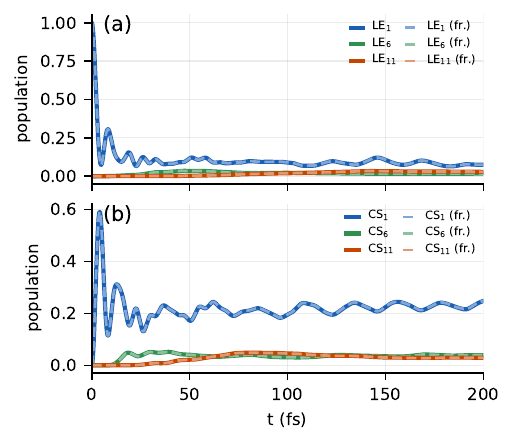}
\vspace{-2em}
\caption{\label{fig:p3ht_dynamics}
Populations of (a) local exciton and (b) charge separated electronic states at various distances from the junction (near: $\mathrm{LE}_1/\mathrm{CS}_1$) (far: $\mathrm{LE}_{11}/\mathrm{CS}_{11}$).  
Fixed $\chi=64$ reference results plotted as solid lines, whereas the adaptive scheme with freezing at the same $\chi=64$ is plotted as dashed lines.}
\end{figure}

The reference system considered here consists of a chain of thirteen oligothiophene units coupled to a fullerene acceptor, leading to a system containing 26 coupled electronic states and 113 vibrational modes\cite{doi:10.1021/acs.jpclett.2c01928,doi:10.1021/ja4093874,doi:10.1021/acs.jctc.4c00751, doi:10.1021/acs.jpclett.5b00336}.
For each oligothiophene molecule with index $n$, this model contains both a local exciton state, $\ket{\mathrm{LE}_n}$, and a charge-separated state, $\ket{\mathrm{CS}_n}$.
The vibrational environment for each state arises from eight intramolecular modes associated with each oligothiophene, a further eight fullerene modes that couple to all charge-separated states, and a reaction-coordinate mode that mediates charge transfer at the donor acceptor interface through a Peierls coupling (see Fig.~\ref{fig:p3ht_pcbm_schematic}).

\begin{figure*}[t!]
\centering
\includegraphics[width=\textwidth]{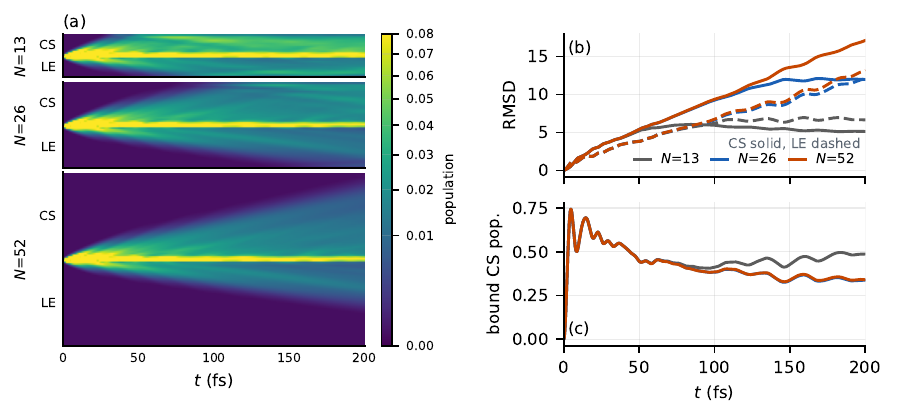}
\vspace{-2em}
\caption{\label{fig:p3ht_finite_size}
Finite size effects on the charge separation dynamics, at $\chi=40$.
(a)~Population dynamics for the 13-, 26- and 52-unit chains. 
(b)~Root-mean-square displacement of each manifold, Eq.~(\ref{eq:manifold-rmsd}),
CS solid and LE dashed.
(c)~The total CS population in the first six units nearest the junction (bound states). All AMTTN simulations used a spawning tolerance of $10^{-4}$.}
\end{figure*}

In Figs.~\ref{fig:p3ht_dynamics}(a) and \ref{fig:p3ht_dynamics}(b), we compare local exciton and charge separated populations obtained using AMTTN with set freezing against the converged fixed $\chi=64$ reference, with good agreement observed throughout the entire propagation. 
Following photoexcitation, there is rapid transfer of population between the local exciton and charge separated manifolds, following which the excitation propagates within the charge separated and local exciton manifold with differing propagation speeds (also illustrated in the top panel of Fig.~\ref{fig:p3ht_finite_size}(a)). 
The adaptive calculations reproduce both the short time population transfer dynamics and the longer time oscillatory features of the reference simulations, with deviations that remain indistinguishable on the scale of the plots.
Importantly, these methods agree both for states located near the interfaces as well as more distant states that are only weakly occupied during the dynamics. 

The $\chi_{\max}=64$ AMTTN calculation in Fig.~\ref{fig:p3ht_dynamics} has a time-averaged population error (Eq.~(\ref{eq:l1-error}) of Appendix~\ref{app:p3ht_convergence}) of $3.2\times10^{-3}$ relative to the fixed $\chi=64$ reference.
This is smaller than the $4.2\times10^{-3}$ error of a fixed $\chi=48$ calculation.
It requires 27 hours on a single CPU core, a factor of $\sim 4$ times faster than the 103 hours required by the fixed $\chi=64$ calculation.
The convergence with respect to $\chi$ and spawning tolerance, and the corresponding runtimes, are presented in Appendix~\ref{app:p3ht_convergence}.

Having established agreement with benchmark calculations, we now employ AMTTN to investigate large systems that are inaccessible to fixed $\chi$ simulations. 
To do so, we consider extended models of 26 and 52 oligothiophene molecules, corresponding to 52 and 104 electronic states, respectively.
In addition to providing a more challenging test for AMTTN, these larger systems are physically important as they delay the onset of artificial reflections in the dynamics due to finite size of the chain.  
For the reference system with $13$ oligothiophene units these finite size effects are observed for $t>50$ fs in the CS state manifold.
Extending the system sizes allows us to probe the charge-separation dynamics for longer times before finite-size effects are observed.
In order to extend the system it is necessary to introduce new values for the charge separated state energies $\epsilon_{\mathrm{CS},n}$ beyond the thirteenth oligothiophene unit.  
Here we take a constant 70~meV for all additional units as the published 13 site model has charge separated state energies that vary between 69.5 and 71.2~meV over units 9-13. 
All remaining parameters are uniform for each unit.

We start by characterising the spatial extent of the exciton dynamics within each electronic manifold. Here we do so by monitoring the root mean square displacement of the population within each electronic manifold separately,
\begin{equation}
R_X(t) = \sqrt{ \frac{\sum_n p^X_n(t)\,(n-n_0)^2}{\sum_n p^X_n(t)} },\qquad X \in \{\mathrm{LE},\mathrm{CS}\},
\label{eq:manifold-rmsd}
\end{equation}
where $p^X_n(t)$ denotes the population of state $\ket{X_n}$ and $n_0$ is the index of the initially excited unit.

In figure~\ref{fig:p3ht_finite_size}(b), we observe identical early time dynamics for the 13-, 26- and 52-unit chains, confirming that the enlarged systems reproduce the benchmark model before the wavepacket reaches the final oligothiophene units. 
However, finite size effects become apparent on intermediate timescales ($t\sim 50$ fs) for the 13 unit simulation, where we begin to see saturation of $R_{\mathrm{CS}}\sim6$ followed by a subsequent decrease, while $R_{\mathrm{LE}}$ saturates at approximately $7$ units near $t\approx 110$~fs. Increasing to 26 units we observe that the timescale for onset of these plateaus increase to $t\sim125$ fs and $t>200$ fs respectively, and finally no evident finite size effects are observed for $N=52$ unit simulations. These effects are also clearly evident in the population dynamics shown in  Fig.~\ref{fig:p3ht_finite_size}(a), in which significant reflections can be observed in the 13 and 26 unit models.
  
Finally, these finite size effects impact estimates of long time populations of bound charge separated excitations (defined as remaining within six sites of the heterojunction\cite{doi:10.1021/acs.jctc.4c00751}).  
 In Fig.~\ref{fig:p3ht_finite_size}(c), we present the bound charge separated state population dynamics. 
 At long times $t>100$ fs, the 13 unit model overestimates the bound state by $\sim 45\%$. In contrast, the 26 and 52 unit chains agree well over the entire time regime, indicating that the bound state population is converged with respect to chain length over these timescales.  The discrepancy in the 13 unit model can therefore be attributed to finite size reflections of the charge separated wavepacket off the boundary.

\begin{figure}[t]
     \centering
     \includegraphics[width=\columnwidth]{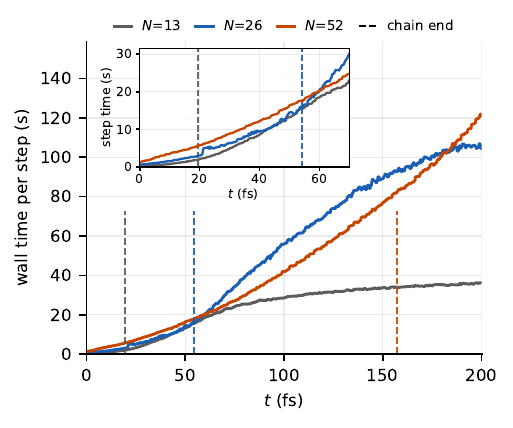}
     \vspace{-2em}
     \caption{Wall time per step obtained from AMTTN calculations for $N=13$, $N=26$, and $N=52$.  The dashed vertical lines indicate the times at which a population of $10^{-6}$ has reached the end of the chain.  The inset shows the short time behaviour.}
     \label{fig:steptime_per_step}
 \end{figure}

 We note that AMTTN provides significant runtime advantages for the 26 and 52 unit models. Here we obtain CPU times of 10.5  hours for $N=13$, 23.7 hours for $N=26$, and 21.3 hours for $N=52$. We observe a reduced runtime for the larger $N=52$ simulation compared to the $N=26$ simulation over the time frame considered, which can be traced to the impact of finite size reflection leading to an increased complexity of the state (and consequently propagation time).  In Fig.~\ref{fig:steptime_per_step}, we show the wall time per simulation step as a function of the simulation time for the simulations of the adaptive bond dimension $N=13$, $N=26$, and $N=52$ against the time at which the population first reaches the end of the chain.  
 For $N=13$ we observe that a population of $10^{-6}$ has reached the final site in the chain by $t\sim 20$~fs, following which the cost per step increases significantly, approaching that of the $N=26$ simulation for $45~\mathrm{fs}\lesssim t\lesssim 50$~fs. 
 Similarly for $N=26$, a population of $10^{-6}$ has reached the final site by $t\sim 54$~fs, following which we observe a significant increase in the runtime per step, with the $N=26$ simulation being more costly than the $N=52$ simulation from $t\sim 60$~fs until $t\sim 180$~fs.
 In contrast, fixed $\chi=40$ simulations take 31.5 CPU hours for $N=13$. 
 For larger $N$ we obtain estimated runtimes by extrapolating the smallest (average) runtime from the first $5$ fs of dynamics of 127 (131) CPU hours for $N=26$ and 854 (1332) CPU hours for $N=52$. 

\section{Conclusion\label{sec:conclusions}}
In this manuscript we have developed the adaptive multiset tree tensor network (AMTTN) framework for quantum dynamics that allocates computational effort where needed.  It adaptively controls bond dimensions throughout the distinct tensor networks through a multiset generalisation of an energy variance based subspace expansion, and adaptively selects the set variables to evolve through the use of a state space restriction scheme. 
The bond dimension adaptation scheme naturally increases bond dimension in unoccupied sets before wavefunction weight reaches the set, while the matrix free Krylov formulation extracts the optimal expansion directions at the same asymptotic cost as standard single site contractions.  

For the one dimensional Holstein model, AMTTN obtains dynamics converged to sub percent accuracy across couplings spanning near ballistic transport to strongly localised polaron formation. Our runtimes are one to two orders of magnitude faster than published fixed bond dimension multiset tensor network simulations \cite{PhysRevLett.123.126601}. Although direct comparisons of runtimes are complicated due to differences in computer hardware, the reduction in computational cost associated with the AMTTN method leads to clear benefits compared to existing methods. 

For the P3HT:PCBM heterojunction model, AMTTN reproduces fixed $\chi$ benchmarks on the standard 13 unit model at reduced computational cost, and allows for extension to chains of 26 and 52 units (up to 104 electronic states with 425 vibrational modes).  
By considering extended chains it is possible to remove boundary reflection finite size artifacts present in the 13 unit dynamics beyond $t\approx 50$ fs.  
While the total charge separated state dynamics is only weakly perturbed by these finite size effects, the dynamics associated with the sets of states partitioned into bound and free charge carriers is not.
The 13 unit chain overestimates the bound fraction by $\simeq 45\%$ at long times due to reflections arising from finite size effects.  
Such effects can be avoided through the use of longer chains.
These calculations demonstrate that the cost of AMTTN is largely determined by the dynamically active region of the wavefunction, with significantly weaker dependence on the lattice size $N$ compared to the fixed $\chi$ simulations.

The methods developed in this manuscript are general and can be applied to any Hamiltonian (or generator of dynamics) expressible in the multiset form of Eqs.~\ref{eq:H1} and \ref{eq:ms-sop}.  
This includes vibronic coupling models, exciton phonon lattices, and hierarchical equations of motion (and related) representations of open quantum systems\cite{TanimuraJPSJ1989,IshizakiJPSJ2005,TanimuraJCP2020,PhysRevLett.120.030402,park2024quasilindbladpseudomodetheoryopen, Lambert2019}.
By extending the range of nonequilibrium vibronic dynamics that can be simulated efficiently on classical computers, AMTTN provides more demanding reference calculations for benchmarking approximate classical methods, including semiclassical approaches, and emerging quantum algorithms\cite{10.1039/d3sc02453a,Motlagh_2025}.

There are several natural extensions of the present work.
The variable freezing approach, applied here to the set variables, could be extended to spatial regions with each tensor network, potentially reducing the residual system size scaling of the present method by removing the cost of sweeping through inactive vibrational degrees of freedom within the problem.  
While here we have considered vibrational environments initially in the vacuum state, the combination of the AMTTN framework with finite temperature and dissipative bath representations~\cite{PhysRevLett.105.050404, 10.1063/1.4866769, 10.1063/1.3490188,Rosenbach_2016} may give access to condensed phase exciton dynamics under experimentally relevant conditions.
Finally, combining the present methods and their finite temperature extensions with memory kernel approaches could provide access to longer timescale dynamics \cite{PhysRevLett.112.110401,10.1063/1.1624830,10.1063/5.0293127}, while combination with space local memory kernel methods \cite{10.1063/5.0249145} may enable accurate simulations of transport in the thermodynamic limit. 
All methods developed in this manuscript are available in the open-source pyTTN package  \url{https://github.com/pyttn/pyTTN}.

    {\noindent}{\bf Code availability}: All methods developed in this paper are available as part of the pyTTN software package \url{https://github.com/pyttn/pyTTN}.  Scripts for performing the simulations presented in this work are provided in the examples folder.

    {\noindent}{\bf Acknowledgements}:     This work was supported by the UK Government Department of Science, Innovation and Technology through the UK National Quantum Technologies Programme  and the Engineering and Physical Sciences Research Council (Grant No. EP/Z53318X/1).  We thank Robin Quade for feedback on the manuscript.

    {\noindent}{\bf Conflict of Interest}:
    The authors have no conflicts to disclose.

    {\noindent}{\bf Data Availability}:
    The data that support the findings of this study are available upon reasonable request. 

\appendix
\section{Comparison with Sparse Wavefunction Methods \label{app:paces}}
To give further context to the performance of AMTTN, in this appendix we compare to the recently reported GPU-accelerated sparse wavefunction method PACES\cite{10.1063/5.0325590}.  
PACES propagates the wavefunction in a time-adapted basis of configurations with size controlled by the truncation number $q_{\rm nom}$ and the neighbour order $m$.
It is designed to make efficient use of a single GPU, and provides an alternative route to the dynamics of vibronic models without the use of tensor network representations.

\begin{figure}[t]
\includegraphics[width=\columnwidth]{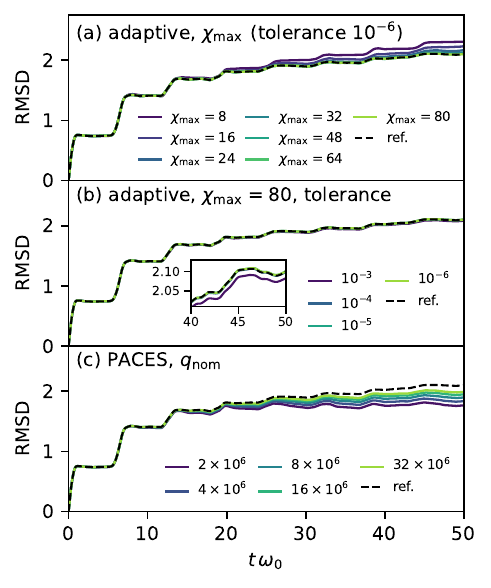}
\vspace{-2em}
\caption{RMSD of the $g=4$ Holstein chain ($N=25$, boson dimension 128) as a function of time.
(a) AMTTN simulations varying the maximum bond dimension $\chi_{\max}=8$-80 at fixed spawning tolerance $10^{-6}$;
(b) AMTTN simulations with $\chi_{\max}=80$ and variable spawning tolerances $10^{-3}$-$10^{-6}$ (inset: the long time behaviour);
(c) PACES with $q_{\rm nom}=2$-$32\times10^{6}$ (with populations renormalised by the norm). The dashed line in each panel is the reference multiset adaptive calculation obtained with $\chi_{\max}=128$ and spawning tolerance $10^{-6}$.
\label{fig:paces}}
\end{figure}

PACES calculations use the publicly available implementation (\url{https://github.com/rkevk/paces}, commit: 90975bc) for the $g=4$ Holstein model with Hamiltonian, initial state, lattice size ($N=25$), time step ($\mathrm{d}t=0.05\,\omega_0^{-1}$) and local boson dimension (128, following Ref.~\citenum{PhysRevLett.123.126601}) all consistent with the main text.  
Following Ref.~\citenum{10.1063/5.0325590} we obtain dynamics using
$m=2$ with an initial neighbour order of 10, and vary $q_{\rm nom}$ from $2\times10^{6}$ up to $32\times10^{6}$, the largest
value that fits within the memory of a single NVIDIA A100 (80\,GB) GPU at $m=2$.
In Fig.~\ref{fig:paces}(a), we present the RMSD convergence obtained using AMTTN at various bond dimensions to the tightly converged reference value $\chi_{\max}=128$, spawning tolerance $10^{-6}$.  
Here we observe systematic convergence with increasing $\chi_{\max}$ with the $\chi_{\max}\geq 48$ results being indistinguishable from the reference on this scale.  
Fig.~\ref{fig:paces}(b) illustrates convergence to the reference with respect to the singular value tolerance obtained at $\chi_{\max}=80$, we observe close agreement for all spawning tolerances considered here with only minor deviations observed at long time for a tolerance of $10^{-3}$.  

\begin{table}[t]
\begin{ruledtabular}
\begin{tabular}{lcccc}
 & $\Delta_{\max}$ & final norm & GPU (h) & CPU (h) \\
\hline
\multicolumn{5}{l}{\textbf{AMTTN, tolerance $10^{-6}$}} \\\hline
\multicolumn{5}{l}{$\chi_{\max}$} \\
8 & $+10.0\%$ & -- & -- & 0.23 \\
16 & $+6.3\%$ & -- & -- & 0.99 \\
24 & $+3.2\%$ & -- & 0.8 & 2.4 \\
32 & $+1.4\%$ & -- & 1.0 & 5.1 \\
48 & $-0.33\%$ & -- & 1.6 & 24 \\
64 & $+0.17\%$ & -- & 2.2 & 38 \\
80 & $+0.088\%$ & -- & 2.7 & 57 \\
\hline
\multicolumn{5}{l}{\textbf{AMTTN, $\chi_{\max}=80$}} \\\hline
\multicolumn{5}{l}{tolerance} \\
$10^{-3}$ & $-0.86\%$ & -- & 0.8 & 3.4 \\
$10^{-4}$ & $-0.12\%$ & -- & 1.7 & 13 \\
$10^{-5}$ & $+0.084\%$ & -- & 2.5 & 37 \\
$10^{-6}$ & $+0.088\%$ & -- & 2.7 & 57 \\
\hline
\multicolumn{5}{l}{\textbf{PACES}} \\\hline
\multicolumn{5}{l}{$q_{\rm nom}$} \\
$2\times10^{6}$ & $-16.0\%$ & 0.943 & 0.12 & -- \\
$4\times10^{6}$ & $-12.6\%$ & 0.960 & 0.23 & -- \\
$8\times10^{6}$ & $-9.7\%$ & 0.973 & 0.45 & -- \\
$16\times10^{6}$ & $-7.2\%$ & 0.982 & 0.92 & -- \\
$32\times10^{6}$ & $-5.2\%$ & 0.989 & 1.9 & -- \\
\end{tabular}
\end{ruledtabular}
\caption{Accuracy and simulation time for the $g=4$ Holstein chain ($N=25$, boson dimension 128) up to $t\omega_0=50$.
$\Delta_{\max}$ is the maximum relative deviation of the RMSD from the reference AMTTN calculation with $\chi_{\max}=128$ and spawning tolerance $10^{-6}$ (4.2 hours on one A100 GPU).
PACES RMSD values are computed from site populations renormalised by the norm at each point in time. 
The reported GPU times were obtained on a single A100 (80~GB), with AMTTN calculations using the hybrid scheme, in which nodes migrate to the GPU as they grow in size. 
AMTTN CPU and GPU simulations with the same parameters agree to better than 0.1\% in RMSD, with tighter agreement at larger $\chi_{\max}$.
\label{tab:paces}}
\end{table}

Finally, in Figure~\ref{fig:paces}(c) we present the same comparison for PACES as a function of $q_{\rm nom}$.
The RMSD approaches the reference systematically as $q_{\rm nom}$ is increased, and the short-time dynamics are reproduced accurately for all values considered.
At longer times, differences are observed between $q_{\rm nom}=16\times10^{6}$ and $32\times10^{6}$, indicating that larger values of $q_{\rm nom}$ are required to converge the long time exciton spreading for this strongly coupled case.

\begin{figure}[t]
\includegraphics[width=0.45\textwidth]{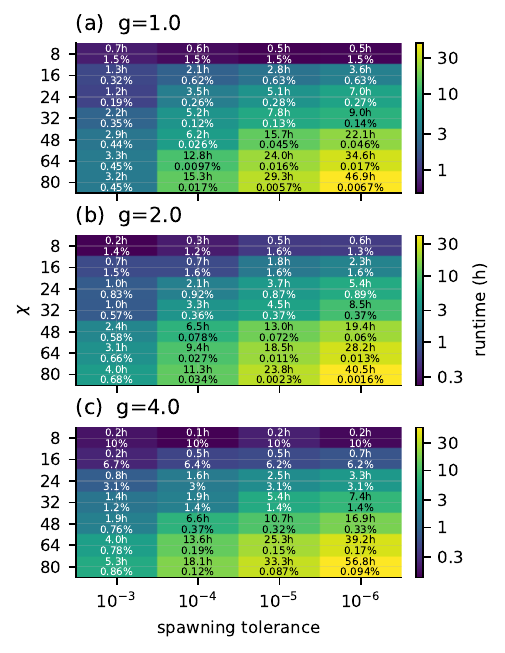}
\caption{\label{fig:holstein_grids}
Single core CPU runtime to $t\omega_0=50$ as a function of the maximum bond dimension $\chi_{\max}$ and the spawning tolerance for (a)~$g=1$, (b)~$g=2$ and (c)~$g=4$.  Each cell gives the runtime and the maximum relative RMSD error against the reference calculation (with spawning tolerance $10^{-6}$ and $\chi_{\max}=128$).  The cell colour shows the runtime in CPU hours associated with each simulation.}
\end{figure}

This can be attributed to the truncation of configurations with small amplitudes.
At the edge of the spreading wavepacket the population on each site is small and typically distributed over many phonon configurations.
As such, these configurations are the first to be removed when the basis is truncated.  
For $q_{\rm nom}=32\times10^{6}$ at $t\omega_0=50$, population for sites at most two hoppings away from the initial excitation agree with the reference to less than 1\%, while those seven to eight sites away are 43\% lower.
As RMSD depends significantly on the outer sites, this leads to an underestimate of the spreading.
In AMTTN each electronic site carries its own tensor network, whose bond dimension adapts to the part of the wavefunction associated with that electronic site. 
This allows for weakly populated sites at the front of the wavepacket to be retained at low cost.

Table~\ref{tab:paces} presents the accuracy and cost of both approaches for this model.
PACES provides an efficient method for obtaining moderate accuracy (convergence to several percent) results for this system.
The key restriction on accuracy is the total accessible GPU memory suggesting that PACES may benefit significantly from newer higher memory devices and multi-GPU implementations.
For the same model, AMTTN calculations reach a deviation of about 1\% in 1.4 hours and about 0.1\% in 2.6 hours on the same GPU (3.4 hours and 13 hours on a single CPU core).
For problems of this type, where the long-time dynamics depends on accurately capturing weakly populated regions of the wavefunction, an adaptive tensor network description therefore provides an effective complement to sparse configuration based approaches.

\section{CPU Runtime Vs Accuracy \label{appendix:cpu_runtime}}

A more detailed picture of the trade off between accuracy and cost can be seen in Fig.~\ref{fig:holstein_grids}.
In this figure we present AMTTN runtimes and accuracies as a function of both the spawning tolerance and the maximum bond dimension.
For small $\chi_{\max}$ the error is set by the maximum allowed bond dimension. 
In these cases, bond dimensions of the set near the exciton saturate rapidly, and consequently reducing the spawning tolerance increases runtime without improving accuracy. 

For larger $\chi_{\max}$ the error is controlled by the spawning tolerance, and reducing the tolerance from $10^{-3}$ to $10^{-4}$ reduces the error significantly (e.g. from $0.45\%$ to $0.019\%$ for $g=1$ and from $0.58\%$ to $0.078\%$ for $g=2$ at $\chi_{\max}=48$).
At a fixed tolerance both the runtime and accuracy become insensitive to $\chi_{\max}$ once it exceeds the bond dimension required by the tolerance.  This is clearly seen for $g=1$ at a tolerance of $10^{-3}$ for $\chi_{\max}\gtrsim48$.
For all $g$ shown here, sub percent accuracy is obtained with a tolerance of $10^{-3}$ and a moderate cap in one to two hours on a single CPU core.

\begin{table}[t]
\begin{ruledtabular}
\begin{tabular}{ccccc}
$g$ & $\chi_{\max}$ & tolerance & time (h) & max($\Delta$RMSD) \\
\hline
1 & 16 & $2\times10^{-3}$ & 0.7 & 0.19\% \\
1.5 & 24 & $3\times10^{-4}$ & 1.6 & 0.41\% \\
2 & 32 & $3\times10^{-4}$ & 1.7 & 0.35\% \\
2.5 & 40 & $5\times10^{-4}$ & 1.8 & 0.34\% \\
3 & 40 & $5\times10^{-4}$ & 1.8 & 0.31\% \\
3.5 & 40 & $5\times10^{-4}$ & 1.9 & 0.54\% \\
4 & 40 & $10^{-3}$ & 1.4 & 0.63\% \\
\end{tabular}
\end{ruledtabular}
\caption{Convergence parameters, runtimes and maximum RMSD of AMTTN simulations of the one dimensional Holstein model completing in under 2 hours on a single CPU core.
RMSD errors are computed against the reference calculation for that system (spawning tolerance $10^{-6}$, $\chi_{\max}=128$).\label{tab:production}}
\end{table}

To identify practical production settings, we additionally performed a sweep over $\chi_{\max} \leq 48$ and spawning tolerances $2\times10^{-3}$-$3\times 10^{-4}$ for all couplings considered in the main text (Table~\ref{tab:production}).  Here we use the same system sizes and local bosonic Hilbert space dimension as considered in Ref.~\citenum{PhysRevLett.123.126601}.  
For all coupling strengths, $g$, AMTTN simulations can obtain sub percent accuracy in under two hours on a single CPU core. For $g \leq 3$ the most accurate run completing within two hours has an error below $0.5\%$. For $g \geq 3.5$, errors below $0.5\%$ require tighter tolerances and 2-3 hours.

\section{Convergence for the 13 unit P3HT:PCBM Model \label{app:p3ht_convergence}}
To monitor the accuracy of the P3HT:PCBM calculations we use the time-averaged error in the electronic state populations,
\begin{equation}
\mathrm{error} = \frac{1}{N_t}\sum_{k=1}^{N_t}\sum_{\alpha=1}^{N_{\rm set}}
\left| p_\alpha(t_k) - p_\alpha^{\rm ref}(t_k) \right|,
\label{eq:l1-error}
\end{equation}
where $p_\alpha(t)=\braket{\Psi^\alpha(t)}{\Psi^\alpha(t)}$ is the population of electronic state $\alpha$.
We use fixed $\chi=64$ simulations for the reference state populations, $p_\alpha^{\rm ref}(t)$.
The outer sum runs over the $N_t$ time points $t_k$ used in the simulation up to $t=200$~fs.

\begin{figure}[t]
\centering
\includegraphics[width=0.5\textwidth]{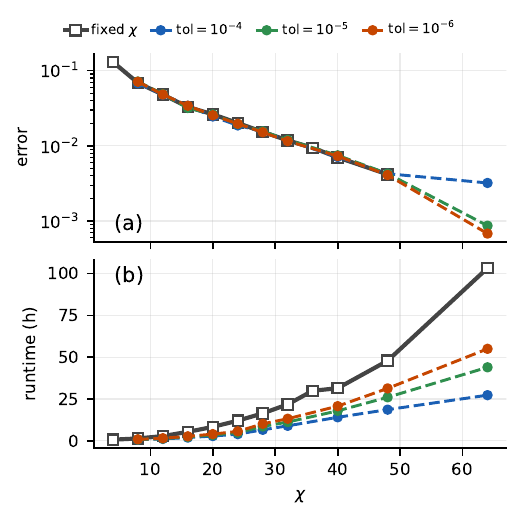}
\caption{\label{fig:p3ht_convergence}
(a)~Population error, Eq.~(\ref{eq:l1-error}), compared to the fixed $\chi=64$ reference and (b)~single core CPU runtime to $t=200$~fs, both as functions of the bond dimension $\chi$ for the 13-unit P3HT:PCBM model.
AMTTN calculations with set freezing (circles) are presented with spawning tolerances $10^{-4}$, $10^{-5}$ and $10^{-6}$ and are compared against fixed bond dimension calculations (squares).}
\end{figure}

A quantitative comparison of the convergence provided in Fig.~\ref{fig:p3ht_convergence}(a), which reports the time averaged population error (across all electronic states) as a function of the maximum bond dimension $\chi_{\rm max}$.
Here we observe systematic convergence of the adaptive calculations towards the fixed bond dimension reference as $\chi$ is increased. Significantly the errors observed for AMTTN are comparable to the errors observed between fixed bond dimension simulations with the same $\chi$ indicating that the adaptive approach is not limiting the accuracy of the simulation for the parameter regimes considered. 

The corresponding CPU times are shown in Fig.~\ref{fig:p3ht_convergence}(b).
Across the entire range of bond dimensions considered, the adaptive propagation scheme uses less CPU time than the  equivalent fixed $\chi$ simulation.  Here we observe considerably smaller reductions in runtime, consistent with rapid spreading of entanglement within this relatively small scale system.  These results demonstrate that neither adaptive bond dimension nor set freezing introduce considerable error beyond that associated with the underlying bond dimension truncation. GPU timings for this model are presented in Appendix~\ref{app:gpu_timings}.

\section{GPU Acceleration \label{app:gpu_timings}}
 AMTTN can benefit from GPU acceleration.
Here we have applied a strategy in which nodes within the tensor network are distributed across the GPU and CPU according to their size, with small nodes handled by the CPU, since for small nodes the GPU offers no advantages. As the bond dimension of small nodes increases to a larger value, these nodes are transferred to the GPU and all future updates of this node are processed on the GPU.
All GPU calculations used a single NVIDIA A100 (80~GB) and a single core of an Intel Xeon Gold 6338 CPU.

\textit{Holstein Model:} For $g=4$ with $\chi_{\max}=80$, the hybrid calculation is faster than the CPU calculation by a factor that grows as the spawning tolerance decreases. 
We find a GPU speed up of 4.3 at $10^{-3}$ (0.8 hours against 3.4 hours), 7.6 at $10^{-4}$ (1.7 hours compared to 13 hours) and 21.1 at $10^{-6}$ (2.7 hours against 56.9 hours).
As the tolerance reduces, we encounter larger bond dimensions at earlier times and across more of the tensor network increasing the percentage of tensors large enough to make use of the larger computational power provided by the A100 GPU.

Additionally, the GPU does not remove the advantage of the adaptive scheme for this model.
With fixed bond dimension throughout the tensor network the GPU simulations take 8.5 hours for $\chi=80$ and 39.9 hours for $\chi=128$.
These calculations keep the GPU fully occupied (97--99\% utilisation), so their cost is limited by the computational costs associated with evaluating the operations on full size tensors at every bond and every timestep.
In contrast, the adaptive calculation requires these costly evaluations only whenever the wavefunction requires it, leading to tightly converged AMTTN runtimes (spawning tolerance $10^{-6}$) of 2.7 hours and 4.2 hours at the same maximum bond dimensions.

\textit{P3HT:PCBM Model}
Running the P3HT:PCBM calculations on a GPU, we find that this model is a less favourable workload than the Holstein chain.
The tensor network contains 224 tensors for each of the 26 electronic states. 
Each tensor has local Hilbert space dimension of 12-24, and bond dimension of at most 64.
Each timestep consists of a very large number of operations (tensor contractions, Krylov operations, and decompositions) on tensors with dimension typically $\mathcal{O}(10)$.
We find that for this model considerable runtime (comparable to the actual operations) is dedicated to kernel launch and synchronisation overhead.
As a result, we find that for $\chi\le 40$ the GPU provides no runtime improvements over a single core CPU, and consequently the mixed CPU/GPU placement strategy implemented within pyTTN favours keeping tensors on the CPU.  
Only once the bond dimension becomes larger do we observe improved GPU times.

For the AMTTN calculations we have runtimes of 5.2 hours and 6.8 hours for $\chi=48$ and $64$, respectively, on a single A100, 3.6 and 4 times faster than on one CPU core.
However, for this model we find that the fixed $\chi=64$ calculation, in which every tensor is at full size throughout has a GPU runtime of 6.4 hours compared to a CPU runtime of 103 hours on one CPU core.  
As such, we see comparable (and here slightly faster) runtimes using fixed $\chi$ for this model.
To reduce launch overhead, operations of the same type on the different electronic states of a node are combined into grouped kernel launches.
For this model, the reduction in work provided by the adaptive scheme comes from operations on small tensors, and reduces the efficiency with which these operations can be batched together.
As the calculation performance is strongly influenced by launch overhead, batching efficiency contributes significantly to the runtime.

\bibliographystyle{aipnum4-2}
\bibliography{references}
\clearpage
\newpage
\onecolumngrid

\end{document}